\documentclass[aps,prl,reprint,superscriptaddress,nofootinbib]{revtex4-2}

\usepackage{amssymb}
\usepackage{amsmath}
\usepackage{braket}

\usepackage{graphicx}
\usepackage[dvipsnames]{xcolor}
\usepackage{xspace}

\usepackage[]{hyperref}
\allowdisplaybreaks

\newcommand{\reffig}[1]{Fig.~\ref{fig:#1}}
\newcommand{\refeq}[1]{Eq.~(\ref{eq:#1})}

\newcommand{\refeqa}[2]{Eqs.~(\ref{eq:#1}) and~(\ref{eq:#2})}

\newcommand{\B}{\mathcal{B}}
\newcommand{\cO}{\mathcal{O}}
\newcommand{\cQ}{\mathcal{Q}}

\renewcommand{\P}{\mathcal{P}}

\newcommand{\T}{\mathcal{T}}

\newcommand{\had}{\text{had}}
\renewcommand{\Im}{\text{Im}}
\newcommand{\OPE}{\text{OPE}}
\newcommand{\lamkin}{\lambda_{\text{kin}}}
\newcommand{\GeV}{\text{GeV}}
\newcommand{\fnl}{f_{\rm nl}^{BK}}

\newcommand{\gho}{\phantom{\hat{}}}

\begin{document}

\title{\vspace*{-0.71cm}
Unitarity bounds with subthreshold and anomalous cuts for \emph{b}-hadron decays}

\author{Abinand Gopal}
\affiliation{Department of Mathematics, UC Davis, Davis, CA 95616, United States}

\author{Nico Gubernari}
\affiliation{DAMTP, University of Cambridge, Wilberforce Road, Cambridge, CB3 0WA, United Kingdom}

\begin{abstract}
We derive a generalisation of the Boyd-Grinstein-Lebed (BGL)  parametrization.
Most form factors (FFs) in $b$-hadron decays exhibit additional branch cuts --- namely subthreshold and anomalous branch cuts --- beyond the ``standard'' unitarity cut.
These additional cuts cannot be adequately accounted for by the BGL parametrization.
For instance, these cuts arise in the FFs for $B\to D^{(*)}$, $B\to K^{(*)}$, and $\Lambda_b\to \Lambda$ processes, which are particularly relevant from a phenomenological standpoint.
We demonstrate how to parametrize such FFs and derive unitarity bounds in the presence of subthreshold and/or anomalous branch cuts.
Our work paves the way for a wide range of new FF analyses based solely on first principles, thereby minimising systematic uncertainties.
\end{abstract}

\maketitle

\section{State of the art}

The thorough study of semileptonic meson decays over the past few decades has led to strong constraints on physics beyond the Standard Model (SM).
These constraints are obtained by combining high-precision measurements and theoretical predictions.
To enhance the indirect searches for New Physics, it is therefore essential to further reduce the theoretical uncertainties.
This is particularly important in view of the LHC Run 3 and Belle II programmes, which will collect an unprecedented amount of data in the coming years, significantly improving the experimental precision of many observables.

For definiteness, we consider semileptonic $B$ meson decays: $B \!\to\! M \ell_1 \ell_2$, where $M$ is a meson and $\ell_{1,2}$ are leptons.
Nevertheless, most of the results derived in this Letter rely on analyticity and unitarity, making them applicable to other hadron decays.

The primary challenge in obtaining accurate predictions for semileptonic decays lies in calculating the hadronic form factors (FFs), which are scalar functions of the momentum transfer squared $q^2$.
We denote a specific FF as $f_\lambda$, where $\lambda$ is a label used to distinguish between different FFs.
FFs are extremely difficult to calculate because they incorporate non-perturbative QCD effects.
As a matter of fact, most FFs are only known at a few isolated $q^2$ points, calculated using lattice QCD~\cite{FLAG} (or Light-Cone Sum Rules~\cite{Colangelo:2000dp,Bharucha:2015bzk,Gubernari:2018wyi,Gubernari:2022hrq}).
It is therefore necessary to extrapolate or interpolate between the known $q^2$ points using a parametrization to obtain predictions of FFs and hence observables in the whole semileptonic region --- i.e. for $q^2 \in[(m_{\ell_1} + m_{\ell_2})^2,s_-]$ with  $s_-\!\equiv\!(m_B-m_M)^2$.
Parametrizations are normally formulated in terms of power series of $q^2$ (or a related variable).
Since in practice only a finite number of parameters can be determined, a method of estimating the truncation error of such series is essential to correctly assess the theoretical uncertainties.

\begin{figure}[t!]
    \centering
    \includegraphics[width=.48\textwidth]{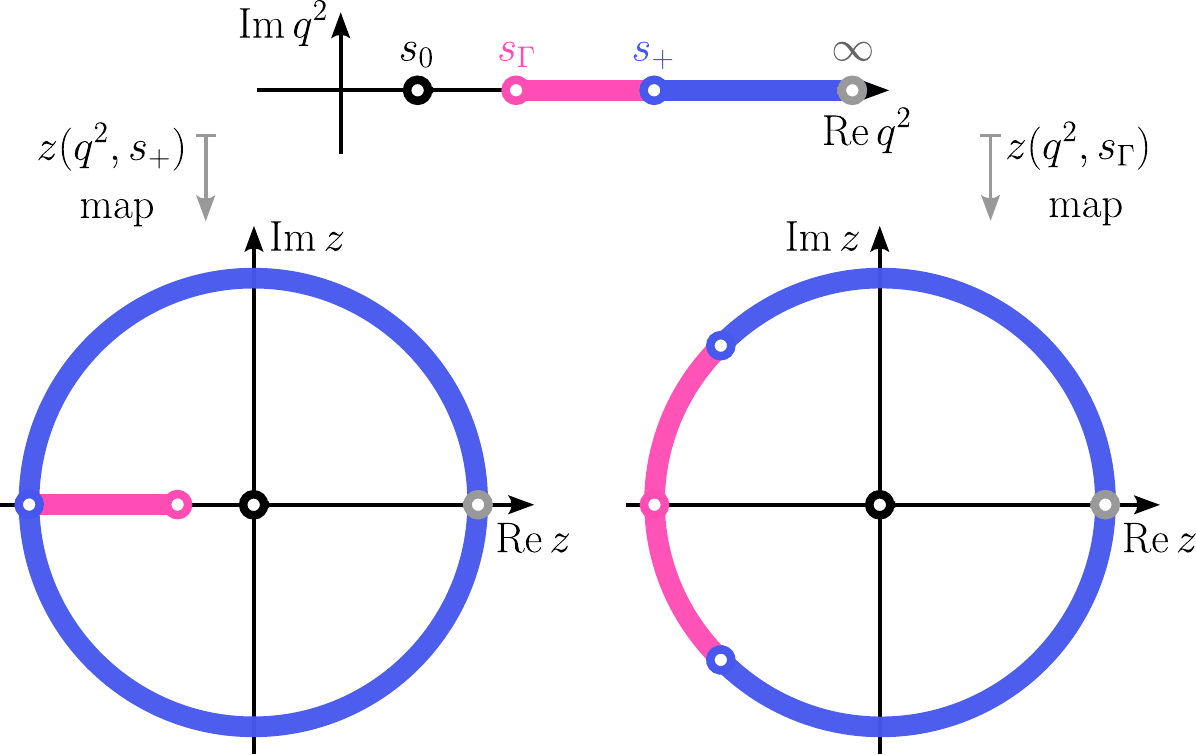}
    \caption{%
        Illustration of the $z(q^2, s_+)$ and $z(q^2, s_\Gamma)$ mappings.
        The magenta line represents the subthreshold cut, while the blue line represents the standard branch cut starting at $s_+$. 
    }
    \label{fig:1}
\end{figure}

It has been shown that the Boyd-Grinstein-Lebed (BGL) parametrization~\cite{Boyd:1994tt,Boyd:1997kz}
\begin{align}
    \label{eq:BGL}
    f_\lambda (q^2)
    =
    \frac{1}{\B_\lambda(q^2) \phi_\lambda(q^2)}\sum_{n=0}^\infty 
    a_{\lambda,n} \, z^n(q^2,s_+)
    \,,
\end{align}
satisfies a unitarity bound (see below).
Here, we have introduced the conformal mapping
\begin{equation}
    \label{eq:zmap}
    z(q^2, s_+) \equiv z(q^2, s_+, s_0) = 
    \frac{
        \sqrt{s_+-q^2}-\sqrt{s_+-s_0^{\phantom{2}}}
    }{
        \sqrt{s_+-q^2}+\sqrt{s_+-s_0^{\phantom{2}}}
    }\,,
\end{equation}
which maps the complex domain $\mathbb{C}\setminus[s_+, \infty)$, with \hbox{$s_+\!\equiv\!(m_B+m_M)^2$}, in the complex $q^2$-plane onto the open unit disk $|z|<1$ in the complex $z$-plane (see \reffig{1}).
The parameter $s_0$ can be freely chosen within the interval $(-\infty, s_+)$, determining the point on the $q^2$-plane that maps to the origin of the $z$-plane.
In \refeq{BGL}, the Blaschke product $\B_\lambda$ has zeros at the isolated poles of $f_\lambda$, while the outer function $\phi_\lambda$ is chosen to be holomorphic with no zeros in $|z|<1$, ensuring that the bound simplifies to the form given below in \refeq{BGL-UB}.
The analytic expressions of $\B_\lambda$ and $\phi_\lambda$ for specific processes can be found in, e.g., Refs.~\cite{Boyd:1997kz,Bharucha:2010im}.

The bound comes from the fact that, using analyticity and unitarity, it is possible to calculate an upper value for the following integral~\cite{Boyd:1997kz}:
\begin{align}
    \label{eq:pre-BGL-UB}
    \frac{1}{\pi}
    \int\limits_{s_+}^\infty \!dq^2 
    \left|\frac{d z(q^2, s_+)}{dq^2}\right|
    \left| \phi_\lambda  f_\lambda \right|^2 
    \!=\!
    \frac{1}{2i\pi}
    \oint_{|z|=1}\!
    \frac{dz}{z}
    \left|
    \phi_\lambda  f_\lambda
    \right|^2
    < 1.
\end{align}
Here, we have omitted the arguments of $\phi_\lambda$ and $f_\lambda$ for brevity and used the fact that $|\B_\lambda|=1$ on the unit circle.
Since the $z^n$ form a complete set of orthonormal polynomials on the unit circle, it follows directly from \refeqa{BGL}{pre-BGL-UB} that
\begin{align}
    \label{eq:BGL-UB}
    \sum_{n=0}^\infty \left|a_{\lambda,n} \right|^2 < 1 \,.
\end{align}
This inequality is called the \emph{unitarity bound}.

The bound \eqref{eq:BGL-UB} provides a systematic method for quantifying the truncation error of the series in \refeq{BGL}, as it is evident that $|z^n| \!\to\! 0:|z| \!<\! 1$ for $n \!\to\! \infty$.
However, the BGL parametrization is only valid for FFs with a branch cut along the real axis from the branch point $s_+$ to infinity.
This is e.g. the case for the FFs in $B \!\to\! \pi \ell\bar\nu$.
When \emph{subthreshold branch cuts} appear —-- i.e., cuts on the real axis starting at $ q^2 \!=\! s_\Gamma \!<\! s_+ $ --— the BGL parametrization no longer applies.
This is e.g. the case for the FFs in $B \!\to\! D^{(*)} \ell\bar\nu$ and the (local and non-local) FFs in $B \!\to\! K^{(*)} \ell^+\ell^-$.
Here, the mapping \eqref{eq:zmap} introduces a branch cut inside the unit disk between $z(s_\Gamma, s_+)$ and $z(s_+, s_+)$ as shown in \reffig{1}.
Consequently, \refeq{pre-BGL-UB} (and thus the bound \eqref{eq:BGL-UB}) cannot be used, since the radius of convergence of the series in \refeq{BGL} is less than one.

Recently, an alternative parametrization has been introduced in Ref.~\cite{Gubernari:2020eft}, offering a partial solution to this issue.
This parametrization uses the mapping $z(q^2, s_\Gamma)$ instead of $z(q^2, s_+)$, with the FFs expanded in terms of polynomials $p_n$ in $z$ that are orthonormal on an arc of the unit circle:\footnote{
    An alternative formulation of this parametrization is given in Ref.~\cite{Flynn:2023qmi}. 
    While it takes a simpler form by using monomials instead of polynomials, the unitarity bound cannot be expressed in a diagonal form,  which prevents an estimation of the truncation error. 
}
\begin{align}
    \label{eq:GVV}    
    f_\lambda (q^2)
    =
    \frac{1}{\B_\lambda(q^2) \phi_\lambda(q^2)}\sum_{n=0}^\infty 
    b_{\lambda,n} \, p_n(z(q^2,s_\Gamma))
    \,.
\end{align}
The explicit expression of $p_{0,1,2}$ is given in Ref.~\cite{Gubernari:2020eft}.
The use of these polynomials is imposed by the fact that in this case the $dz$ integral in \refeq{pre-BGL-UB} does not extend over the entire circle, but only over the arc that goes from $e^{-i\arg\left(z(s_+,s_\Gamma)\right)}$ to $e^{+i\arg\left(z(s_+,s_\Gamma)\right)}$ due to the different mapping (blue arc in the right panel of \reffig{1}).
%
%
Although the coefficients $b_{\lambda,n}$ satisfy a unitarity bound analogous to that in \refeq{BGL-UB}, the absolute value of $p_n$ diverges exponentially as $n \to \infty$ for certain values of $z$ within the unit disk.
Therefore, the parametrization \eqref{eq:GVV} is \emph{not} unitarity-bounded, as even a very small coefficient $b_{\lambda,n}$ can produce a large contribution when $n$ is large.
In addition, since FFs have a branch point at $s_+$, the coefficients $b_{\lambda,n}$
will only decay algebraically with respect to $n$ (i.e., there exists $C > 0$ such that $|b_{\lambda,n}| = o(n^{-C})$ as $n \to \infty$). 
Thus, \refeq{GVV} will not in general converge for $q^2$ such that $q^2 \not\in [s_{\Gamma},s_+]$. 
For a detailed treatment of the convergence of orthogonal polynomials, we refer the reader to Ref.~\cite{Trefethen:2019atap}. 
%

Another challenge in FF parametrizations comes from \emph{anomalous branch cuts}. 
These cuts do not correspond to any physical particle production channel and appear in certain (non-local) FFs in rare semileptonic decays -- see Ref.~\cite{Mutke:2024tww} for a recent discussion.
Notably, these cuts can extend into the complex plane rather than being confined to the real axis.
To date, there is no parametrization that both satisfies a unitarity bound and accounts for anomalous branch cuts.

In the remainder of this Letter, we derive for the first time bounded parametrizations that account for subthreshold and anomalous branch cuts.

\section{Subthreshold branch cuts}

To derive a bounded parametrization for FFs with a subthreshold branch cut, we construct a correlator of the form
\begin{align}
\label{eq:Pi-general}
    \!\!\!\Pi^J(q^2)
    \equiv
    i \,\P_{\mu\nu}^J
    \!\int\!  d^4 x\, e^{i q\cdot x}\bra{0}
    \T\left\lbrace 
        \cO^\mu(x) 
        \cO^{\nu,\dagger}(0) 
    \right\rbrace  
    \ket{0}.
\end{align}
Here, $\cO^\mu$ denotes an operator that can be either local, such as a quark current, or non-local, such as the time-ordered product of an effective four-quark operator and the electromagnetic current, as in Ref.~\cite{Gubernari:2020eft}. 
The projectors are defined as
\begin{align}
    \label{eq:corr}
    \P_{\mu\nu}^0 & \equiv \frac{q_\mu q_\nu}{q^2}\,, 
    &
    \P_{\mu\nu}^1 & \equiv 
    \frac{1}{d-1}
    \left(\frac{q_\mu q_\nu}{q^2} - g_{\mu\nu}\right),
\end{align}
where $d$ is the number of spacetime dimensions.
It is well known that a correlator of the form of \refeq{corr} satisfies a subtracted dispersion relation~\cite{Barton:1965,Colangelo:2000dp}
\begin{align}
    \label{eq:chi}
    \chi^J(Q^2,l)
    \!\equiv\! 
    \frac{1}{l!} \!
    \left[\frac{\partial}{\partial q^2}\right]^l \!
    \Pi^J(q^2)
    \bigg|_{\scalebox{0.6}{$q^2\!=\!Q^2$}}  
    \!=\!
    \frac{1}{\pi} \!
    \int\limits_{0}^\infty\! dq^2
    \frac{\Im\, \Pi^J(q^2)}{(q^2-Q^2)^{l+1}}
    ,
\end{align}
since $\Pi^J$ has singularities exclusively along the positive real axis.
Here $Q^2$ denotes the subtraction point, and $l$ represents the number of subtractions, chosen such that $\chi^J$ remains finite.
For real $Q^2$ values  significantly below the first threshold of $\Pi^J$, the function $\chi^J$ can be systematically calculated using an operator product expansion (OPE) and perturbation theory.
$\chi_\OPE^J$ is known with percent-level accuracy for all phenomenologically relevant cases.

The imaginary part of $\Pi^J$ can be determined using unitarity, i.e. by inserting a complete set of states in \refeq{corr}~\cite{Boyd:1997kz,Caprini:1997mu}.
For the sake of concreteness, we consider the operator $\cO^\mu \!=\! \bar{s} \gamma^\mu b$ and $J\!=\!1$ in the following derivation, although it can easily be generalised to other operators and to $J\!=\!0$.
In this case, the two-particle contribution of the $\bar{B}K$ states to $\Im\,\Pi^J$ reads\footnote{
    The mathematical steps to derive this equation are standard and are omitted here for brevity. Detailed derivations can be found, for example, in Refs.~\cite{Boyd:1997kz,Bharucha:2015bzk,Gubernari:2020zil}.
}
\begin{align}
    \label{eq:ImPi1}
    \Im\,\Pi_{\had}^{1,bs} (q^2)= 
    \frac{|\lamkin|^{\frac32}}{24 \pi(q^2)^2}
        \, |f_+^{BK}(q^2)|^2 \,\theta(q^2-s_+)
     +\dots\,,
\end{align}
where $\lamkin\!\equiv\! \lambda(q^2,m_B^2,m_K^2)$ is the Källén function and the isospin limit has been used.
The ellipsis denotes the contribution of all the other states.
The contribution of the one-particle state, (i.e, $\bar{B}_s^*$) or other two-particle states (e.g. $\bar{B}K^{(*)}$, $\bar{B}_s\phi$, and $\Lambda_b\Lambda$) can easily be added to the r.h.s. of \refeq{ImPi1}~\cite{Bharucha:2010im,Blake:2022vfl,Gubernari:2023puw}. 
The FF $f_+^{BK}$ is defined as~\cite{Gubernari:2018wyi}
\begin{align}
    \P_{\mu\nu}^1\!
    \bra{\bar{K}(k)} \bar{s}\gamma^\mu b \ket{\bar{B}(q+k)}
    \!=\!
    \frac{2}{3}\!
    \left(
         k_\nu \!-\! \frac{q\cdot k}{q^2} q_\nu
    \right)
    f_+^{BK}(q^2)
    .
\end{align}

By equating the OPE result and the hadronic representation \eqref{eq:ImPi1} of $\chi^{1,bs}$ we obtain
\begin{align}
    \label{eq:match1}
    \chi_\OPE^{1,bs}(0,l)
    >
    \int\limits_{s_+}^\infty\! dq^2
    \frac{|\lamkin|^{\frac32}}{24 \pi^2(q^2)^{l+3}}
        \, |f_+^{BK}(q^2)|^2
    \,,
\end{align}
where $s_+\!\equiv\!(m_B+m_K)^2$.
For simplicity, we have set $Q^2=0$, which is the most common choice in the literature~\cite{Boyd:1997kz,Caprini:1997mu}.
Hereafter we omit the first argument of $\chi^J$, namely $\chi^J(0,l)\equiv \chi^J(l)$.
It has been shown that the minimum number of subtractions required to obtain a finite $\chi^{1,bs}$ is $l=2$~\cite{Boyd:1997kz}.
Note that neglecting the additional contributions denoted by the ellipsis in \refeq{ImPi1} and using an inequality sign in \refeq{match1} is justified, as these contributions are positive \hbox{definite~\cite{Boyd:1997kz,Gubernari:2023puw}.}

The domain of analyticity for $f_+^{BK}$ is $\mathbb{C}\setminus\{m_{B_s^*}^2\} \setminus [s_\Gamma, \infty)$. 
In this case $s_\Gamma \!\equiv\! (m_{B_s}+m_\pi)$, since $\bar{B}_s\pi_0$ is the lightest multiparticle state that goes on-shell~\cite{Gubernari:2023puw}.
In addition there is a simple pole at $q^2=m_{B_s^*}^2$.
In the standard approach, the branch cut between $s_+$ and $s_\Gamma$ is typically ignored, allowing the direct application of the procedure outlined in the previous section, as \refeq{match1} can be easily recast in the form of \refeq{pre-BGL-UB} to yield \refeq{BGL-UB}~\cite{Boyd:1997kz,Bharucha:2010im}.
Although it can be argued that these subthreshold cuts give only a minor contribution in certain cases, neglecting them is generally unjustified.
Indeed, expanding a FF using a power series in a non-analytic region introduces unknown systematic uncertainties.
%

To overcome these limitations we add the same (positive) term on both sides of \refeq{match1} and set $l=2$:
\begin{align}
    \label{eq:add-term}
    \Delta\chi^{1,bs}(2)
    \equiv\!
    \int\limits_{s_\Gamma}^{s_+}\! dq^2
    \frac{|\lamkin|^{\frac32}}{24 \pi^2(q^2)^{5}}
        \, |f_+^{BK}(q^2)|^2 .
\end{align}
This term is visually represented by the magenta arc in the right panel of \reffig{1}, whereas the r.h.s. of \refeq{match1} corresponds to the blue arc.
This yields
\begin{align}
    \label{eq:match-new1}
    \tilde\chi^{1,bs}(2)
    >\!
    \!\int\limits_{s_\Gamma}^\infty\! dq^2
    \frac{|\lamkin|^{\frac32}}{24 \pi^2(q^2)^{5}}
        \, |f_+^{BK}(q^2)|^2 ,
\end{align}
where $\tilde\chi^{1,bs} \equiv \chi_\OPE^{1,bs} + \Delta\chi^{1,bs}$.
To estimate $\Delta\chi^{1,bs}$ one can approximate $f_+^{BK}$ using its scaling large $q^2$ calculated in perturbative QCD as~\cite{Lepage:1980fj,Akhoury:1994tnu}
\begin{equation}
    |f_+^{BK}(q^2)|^2 \simeq 
    K 
    \left(\frac{s_\Gamma}{q^2}\right)^2
    \,.
\end{equation}
Following Ref.~\cite{Becher:2005bg}, it is possible to show that the constant $K$ is expected to be smaller than one (see also Ref.~\cite{Boyd:1995sq,Caprini:1995wq,Beneke:2000ry}). 
Even assuming $K \!\sim\! \cO(10^2)$ the contribution of  $\Delta\chi^{1,bs}$ remains less than one per cent of $\chi_\OPE^{1,bs}$, which is significantly smaller than the uncertainty of $\chi_\OPE^{1,bs}$ itself.
The discussion about the appropriate approximation for $f_+^{BK}$ is therefore irrelevant, allowing us to simply set $\tilde{\chi}^{1,bs} = \chi_{\OPE}^{1,bs}$.
This is not surprising, as the interval $[s_\Gamma, s_+]$ is relatively minuscule, making its contribution naturally negligible.

Based on the considerations above, we propose to parametrize $f_+^{BK}$ as
\begin{align}
    \label{eq:GGp}
    f_+^{BK} (q^2)
    =
    \frac{1}{\B_+(q^2)\phi_+(q^2)}\sum_{n=0}^\infty 
    c_{+,n}\, z^n(q^2,s_\Gamma)
    \,,
\end{align}
where $\B_+(q^2) = z(q^2,s_\Gamma, m_{B_s^*}^2)$ and 
\begin{equation}
\begin{aligned}
    \label{eq:phip}
    \!\!\phi_+ (q^2) 
    & = 
    \frac{
        1
    }{
        \sqrt{24^{\phantom{1}}\!\! \pi\, \tilde\chi^{1,bs}(2)}
    }
    \,
    \frac{
        \left( s_+ - q^2 \right)^{\frac34}
    }{
        \sqrt{-d z(q^2,s_\Gamma)/dq^2}
    }
    \!\!
    \\ & \times 
    \frac{
        \left( \sqrt{s_\Gamma - q^2} + \sqrt{s_\Gamma - s_-\gho} \right)^{\frac 3 2}
    }{
        \left( \sqrt{s_\Gamma - q^2} + \sqrt{s_\Gamma\gho} \right)^{5}
    }
    .\!
\end{aligned}
\end{equation}
Note that this outer function depends on both thresholds: $s_\Gamma$ and $s_+$.
Using the mapping $z(q^2, s_\Gamma)$ and \refeq{phip}, \refeq{match-new1} can be written as
\begin{align}
    \frac{1}{2i\pi}
    \oint_{|z|=1}\!
    \frac{dz}{z}
    \left|
    \phi_+(q^2)  f_+^{BK}(q^2)
    \right|_{\scalebox{0.6}{$q^2=q^2(z,s_\Gamma)$}}^2
    < 1\,.
\end{align}
which has exactly the same form of \refeq{pre-BGL-UB} and hence, using \refeq{GGp},
\begin{align}
    \label{eq:GGp-UB}
    \sum_{n=0}^\infty \left|c_{+,n} \right|^2 < 1 \,.
\end{align}
We have therefore derived the first model-independent and unitarity-bounded parametrization for $f_+^{BK}$.
This parametrization can be extended to other FFs and transitions.
While its extension is generally straightforward --- requiring only the calculation of outer functions and the use of appropriate Blaschke factors --- two specific cases require further consideration.

The first is when the term $\Delta\chi^J$ is not negligible. 
This can potentially occur when $(s_+ - s_\Gamma)/s_\Gamma\sim\cO(1)$, a condition that is satisfied by certain non-local FFs in rare semileptonic decays.\footnote{
    The quantity $(s_+-s_\Gamma)/s_\Gamma$ can be relatively large even for (local) FFs of decays involving excited mesons or baryons. 
    In such cases, a careful estimate of $\Delta \chi^J$ is necessary. 
    Importantly, a technique to suppress the contribution of $\Delta \chi^J$ is discussed at the end of this section.
}
These FFs are discussed in the next section.

The second is when the considered FF has a pole between $s_\Gamma$ and $s_+$.
This happens for instance in the FF $f_0^{BK}(q^2)$, defined as
\begin{align}
    \!\P_{\mu\nu}^1\!
    \bra{\bar{K}(k)} \bar{s}\gamma^\mu b \ket{\bar{B}(q+k)}
    =
    q_\nu
    \frac{\sqrt{s_-s_+\gho }}{q^2}
    f_0^{BK}(q^2)
    \,,
\end{align}
which has a simple pole at $q^2\!=\!m_{B_{s0}}^2\!=\!(5.711\,\GeV)^2$~\cite{Lang:2015hza}, and hence $s_\Gamma < m_{B_{s0}}^2 <s_+$.
The minimal number of subtractions for $\chi^{0,bs}$ is $l=1$~\cite{Boyd:1997kz}.
Using \refeq{chi} and setting $Q^2=0$ once again, we construct the following equation:
\begin{equation}
\begin{aligned}
    \label{eq:chi-new0}
    \chi_\Sigma^{0,bs}
    & \equiv
    m_{B_{s0}}^4\chi_\OPE^{0,bs}(3)
    - 2  m_{B_{s0}}^2\chi_\OPE^{0,bs}(2)
    + \chi_\OPE^{0,bs}(1)
    \\
    & =
    \frac{1}{\pi} \!
    \int\limits_{0}^\infty\! dq^2
    (q^2-m_{B_{s0}}^2)^2
    \frac{\Im\, \Pi^{0,bs}(q^2)}{(q^2)^{4}}
    \,.
\end{aligned}
\end{equation}
By substituting $\Im\, \Pi^{0,bs}$ with the contribution from $\bar{B}K$ states (see, e.g., Refs.~\cite{Boyd:1997kz,Bharucha:2015bzk,Gubernari:2020zil}) and following a procedure similar to that used to derive \refeq{match-new1}, we obtain
\begin{multline}
    \label{eq:match-new0-b}
    \tilde\chi^{0,bs}
    \!>\!
    \int\limits_{s_\Gamma}^\infty\! dq^2
    \frac{s_-s_+|\lamkin|^{\frac12}}{8 \pi^2(q^2)^{6}}
    (q^2-m_{B_{s0}}^2)^2
        \, |f_0^{BK}(q^2)|^2 .
\end{multline}
Here $\tilde\chi^{0,bs} \equiv \chi_\Sigma^{0,bs} + \Delta\chi^{0,bs}$ and $\Delta\chi^{0,bs}$ is defined as the integral in the above equation, evaluated over the range from $s_\Gamma$ to $s_+$.
In this way the pole at $q^2=m_{B_{s0}}^2$ in the integrals has been manifestly removed.
We therefore parametrize $f_0^{BK}$ as
\begin{align}
    \label{eq:GG0}
    f_0^{BK} (q^2)
    =
    \frac{1}{\phi_0(q^2)}\sum_{n=0}^\infty 
    c_{0,n}\, z^n(q^2,s_\Gamma)
    \,,
\end{align}
The outer function reads
\begin{equation}
\begin{aligned}
    \label{eq:phi0}
    \phi_0 (q^2) 
    & = 
    \sqrt{
        \frac{
            s_-s_+
        }{
            8^{\phantom{1}}\!\! \pi\, \tilde\chi^{0,bs}
        } 
    }
    \,
    \frac{\left( s_+ - q^2 \right)^{\frac14}}{\sqrt{-d z(q^2,s_\Gamma)/dq^2}}
    \\ & \times 
    \frac{
        \left( \sqrt{s_\Gamma - q^2} + \sqrt{s_\Gamma - s_-\gho} \right)^{\frac12}
    }{
        \left( \sqrt{s_\Gamma - q^2} + \sqrt{s_\Gamma\gho} \right)^{6}
    }
    \left(m_{B_{s0}}^2 - q^2 \right)
    .
\end{aligned}
\end{equation}
Note that no Blaschke factor is needed in this case, as there are no poles below $s_\Gamma$. 
Similar to the case of $f_+^{BK}$, we can derive a unitarity bound using \refeqa{match-new0-b}{GG0}:
\begin{align}
    \label{eq:GG0-UB}
    \sum_{n=0}^\infty \left|c_{0,n} \right|^2 < 1 \,.
\end{align}

The procedure presented in this section to derive unitarity-bounded parametrizations for FFs with subthreshold cuts is general and can therefore be applied to other hadronic decays.
The only step that requires special attention is the evaluation of $\Delta\chi^J$, which involves calculating (or at least estimating an upper bound for) the weighted integral of the squared modulus of the FF along the subthreshold cut.
However, the contribution of $\Delta\chi^J$ is typically negligible in most applications, as subthreshold cuts are usually short.
It is also possible to perform additional subtractions to further suppress the contribution of $\Delta\chi^J$.
This can be accomplished by applying the same procedure we used to remove the pole at $m_{B_{s0}}^2$ in $f_0^{BK}$, replacing $m_{B_{s0}}^2$ with $(s_\Gamma + s_+)/2$, i.e. the midpoint between $s_\Gamma$ and $s_+$.

We conclude this section by noting that, in addition to the approach of Refs.~\cite{Gubernari:2020eft,Flynn:2023qmi} described above, another method for dealing with subthreshold cuts was proposed in Refs.~\cite{Boyd:1995sq,Caprini:1995wq}.
This method uses the mapping $z(q^2,s_+)$ and relies on the cancellation of the subthreshold branch cut within the unit disk.
However, the precise functional form of the FFs along the cut is not known, making it impossible to exactly cancel the cuts within the unit disk.
As a result, while the method of Refs.~\cite{Boyd:1995sq, Caprini:1995wq} can provide a good approximation in many cases, it is generally not reliable and model dependent.

\section{Anomalous branch cuts}

Non-local FFs, which parametrize the hadronic matrix elements of non-local operators~\cite{Khodjamirian:2010vf,Khodjamirian:2012rm,Gubernari:2020eft,Ciuchini:2022wbq,Gubernari:2022hxn,Feldmann:2023plv,Isidori:2024lng}, are known to potentially exhibit anomalous branch cuts~\cite{Lucha:2006vc,Colangelo:2015ama,Mutke:2024tww}.
Anomalous branch cuts, unlike the unitarity cuts discussed in the previous section, are not restricted to lie on the real axis.

To date, there is no unitary-bounded para\-metri\-zation that can account for anomalous branch cuts.
Given the critical role of non-local FFs in predicting observables for rare semileptonic decays~\cite{Khodjamirian:2010vf,Khodjamirian:2012rm,Ciuchini:2022wbq,Gubernari:2022hxn}, the development of such a parametrization is of utmost importance.

For definiteness we consider the non-local FF appearing in $B \!\to\! K \ell^+\ell^-$ decays
\begin{multline}
    \!\!\!\!\!
    i\!\int\! d^4x\, e^{iq\cdot x}
    \bra{\bar{K}(k)} 
    \T\left\{ 
        \bar{c}\gamma_\mu c(x), 
        \cQ(0)
    \right\} 
    \ket{\bar{B}(q+k)} 
    \\*
    \label{eq:def:nonlocBK}
    = 
    2\, m_B^2
    \left(
         k_\mu \!-\! \frac{q\cdot k}{q^2} q_\mu
    \right)
    \, \fnl (q^2) \,,
\end{multline} 
where $\cQ\equiv C_1\cQ_1 +C_2\cQ_2$, with $C_{1,2}$ denoting the Wilson coefficients of the four-quark effective operators
$
\cQ_{2(1)} \!\equiv\! 
\big[\bar{s}_L\gamma^\mu (T^a) q_L  \big]
\big[\bar{q}_L \gamma_\mu (T_a) b_L  \big]$~\cite{Buchalla:1995vs}.
The FF $\fnl$ serves as an excellent example not only because it exhibits anomalous cuts in the complex plane but also due to its significant phenomenological relevance, especially in light of the tensions between measurements and predictions in $B \!\to\! K \mu^+ \mu^-$ decays~\cite{LHCb:2014cxe,CMS:2024syx,Gubernari:2022hxn,Parrott:2022zte,Alguero:2023jeh}.

\begin{figure}[t!]
    \centering
    \includegraphics[width=.48\textwidth]{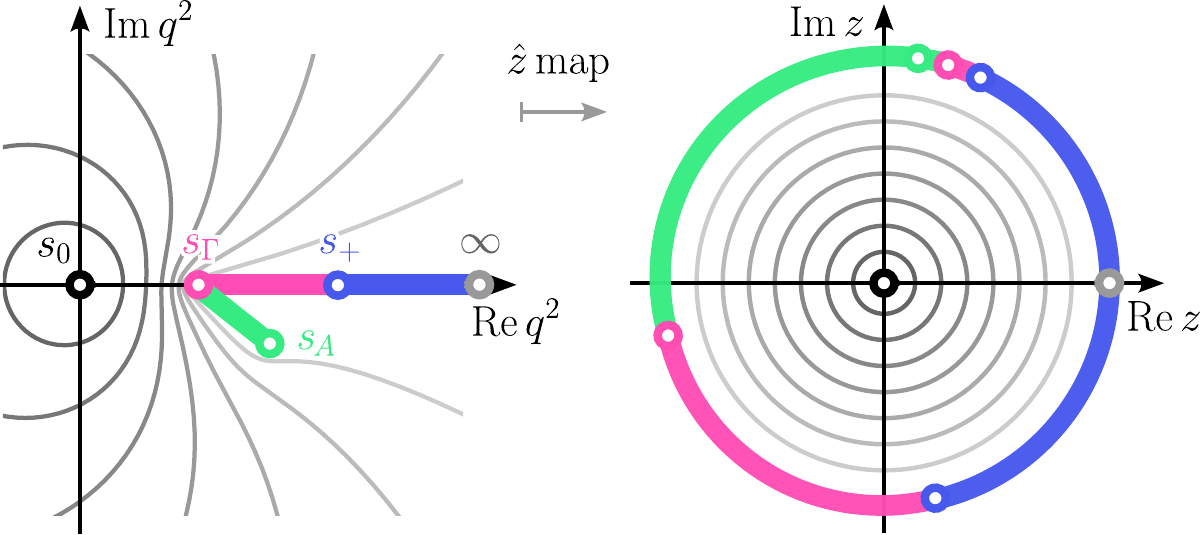}
    \caption{%
        Illustration of the conformal mapping $\hat z$ with \hbox{$s_0\!=\!0$}. 
        We use the same colour scheme as in \reffig{1} with the anomalous branch cut represented in green.  
    }
    \vspace*{-0.2cm}
    \label{fig:2}
\end{figure}

The rescattering process $B \!\to\! D D_s^* \!\to\! K \ell^+ \ell^-$ induces an anomalous branch cut in $\fnl$ between the points $s_\Gamma=4m_D^2$ and $q^2 \!=\! s_A \!\equiv\! 24.1 - 3.5i$, as shown in Ref.~\cite{Mutke:2024tww}.\footnote{
    To simplify the discussion, we omit the additional anomalous cuts generated by other rescattering processes. While their inclusion would require additional effort, it can be done systematically using the method presented here.
}
In addition, $\fnl$ presents a unitarity cut along the real axis starting at $s_\Gamma$.
This domain is illustrated in the left panel of \reffig{2}.

The first step in deriving a parametrization for this case
is to identify a mapping $\hat{z}(q^2)\!\equiv\! z(q^2,s_A,s_\Gamma,s_0)$ that conformally maps the domain 
\setlength{\abovedisplayskip}{7pt}
\setlength{\belowdisplayskip}{9pt}
\begin{equation}
    \Omega = \mathbb{C} \backslash ([s_{\Gamma},\infty] \cup \{ (1-t) s_A + t s_\Gamma:  t \in [0,1] \})
\end{equation}
to the unit disk $D = \{ z : |z| < 1\}$ such that $\hat{z}(s_0)=0$.
The existence of such a mapping is ensured by the Riemann Mapping Theorem.
To compute it we first compute the conformal mapping $g$ from $D$ to $\Omega$ such that
$g(-1)=s_A$, $g(-i)=s_\Gamma$, $g(1)=\infty$. 
We note that since $\Omega$ is a polygon with vertices $w_1 = s_\Gamma$, $w_2 = s_A$, $w_3 = s_\Gamma$, $w_4 = \infty$, with corresponding angles $\phi_1= -\arg(s_A-s_\Gamma)$, $\phi_2=2 \pi$, $\phi_3 = 2\pi-\phi_1$, $\phi_4=-2\pi$, $g$  may be computed using the Schwarz--Christoffel formula:
\begin{equation}
    g(z) = A + C
    \int^{z}_0 d\zeta
    \prod_{k=1}^4
    \left(
        1 - \frac{\zeta}{z_k}
    \right)^{\phi_k/\pi-1},
\end{equation}
where $z_2=-1$, $z_3 = -i$, and $z_4 = 1$; $z_1$, $A$, and $C$ are parameters
that must be determined.
We proceed with the same approach as the Schwarz--Christoffel Toolbox
\cite{Driscoll:sctoolbox} which is described in detail in
\cite{Driscoll_Trefethen_2002}.
We first determine $z_1$ with $|z_1|=1$ such that $g$ satisfies $\int_{z_1}^{z_2} |d\zeta | |g'(\zeta)|=\int_{z_2}^{z_3} |d\zeta| |g'(\zeta)|$ using a
quasi-Newton method.
We then fix $A$ and $C$ by enforcing that $g(z_1)=s_\Gamma$ and $g(z_2)=s_A$.
After $g$ is determined, we can evaluate $g^{-1}$ in $\Omega$ using a Newton iteration.
We now compute $z_0 = g^{-1}(s_0)$ and note that the M\"obius transform $h(z) = (z+z_0 e^{i \theta})/(e^{i \theta} + \overline{z_0} z)$ with $\theta= \arg((1-\overline{z_0})/(1-z_0))$ maps $D$ to $D$, and satisfies $h(0) = z_0$  and $h(1)=1$. 
Thus, $\hat{z}^{-1} = g \circ h$ and $\hat{z}$ can be computed using a Newton iteration on $\hat{z}^{-1}$.
The Python code for computing the conformal mapping with a single branch cut, along with its inverse and derivatives, is provided as 
an ancillary file in the arXiv version of this Letter.

Now that the mapping has been obtained, it is possible to follow the procedure presented in the previous section to obtain a parametrization for $\fnl$.
In fact, from a conceptual point of view, anomalous cuts can be treated as subthreshold cuts.
The OPE calculation and the imaginary part in terms of $\bar{B}K$ states of the corresponding correlator can be found in Ref.~\cite{Gubernari:2020eft}.
The derivation of the outer function is standard and is not shown here.  
The only point requiring further clarification is the calculation of $\Delta \chi^J$, which is particularly challenging in this case and beyond the scope of this work.
Nevertheless, an upper bound for $\Delta\chi^J$ can be obtained by using the fact that it can be decomposed as $\fnl=f_{\rm unit}^{BK}+f_{\rm an}^{BK}$.
The part containing only the unitarity cut $f_{\rm unit}$ can be computed with the local OPE of Refs.~\cite{Grinstein:2004vb,Beylich:2011aq}, while the part containing the anomalous cut $f_{\rm an}$ can be estimated using e.g. the approach of Ref.~\cite{Mutke:2024tww}.
\vspace*{-0.17cm}

\section{Conclusion and outlook}
\vspace*{-0.17cm}

We have presented a systematic procedure for deriving hadronic form factor (FF) parametrizations that satisfy unitarity bounds in the presence of subthreshold and/or anomalous branch cuts.
While FFs with subthreshold cuts are relatively common in hadron decays, anomalous cuts are a distinctive feature of FFs arising from more complicated (non-local) operators.

Our parametrization can be regarded as an extension of the BGL parametrization~\cite{Boyd:1994tt,Boyd:1997kz}, which is not valid when the aforementioned cuts are present.
Moreover, our approach supersedes previous attempts in the literature to account for subthreshold cuts, as it allows a rigorous estimation of the truncation error and does not rely on numerical cancellation of singularities~\cite{Boyd:1995sq,Caprini:1995wq,Gubernari:2020eft,Flynn:2023qmi}.

This work is important not only from a conceptual and theoretical perspective, but also from a phenomenological one.
In fact, the method presented here paves the way for new phenomenological analyses that can rely on first-principles constraints, i.e. the unitarity bounds, to estimate the truncation error.
These constraints will contribute significantly to solving pressing problems such as the tensions in $B\!\to\! K^{(*)} \mu^+\mu^-$ decays~\cite{LHCb:2014cxe,CMS:2024syx,Gubernari:2022hxn,Parrott:2022zte,Alguero:2023jeh}.

\section{Acknowledgements}

This work was partially supported by STFC HEP Consolidated grants
ST/T000694/1 and ST/X000664/1.
NG thanks the Cambridge Pheno Working Group for fruitful discussions.
NG also thanks M. Reboud, D. van Dyk, S. Mutke, M. Hoferichter, and B. Kubis for helpful discussions and comments on the manuscript.

\bibliography{references.bib} 

\end{document}